\documentclass[prb,showpacs,12pt]{revtex4}
\begin{document}
\title{Relationship between Electronic and Geometric Structures of
the O/Cu(001) System}
\author{Sergey Stolbov}
\author{Talat S. Rahman}
\affiliation{Department of Physics, Cardwell Hall,
Kansas State University, Manhattan, KS 66506}

\begin{abstract}
The electronic structure of the
$(2\sqrt{2}\times\sqrt{2})R45^{\circ}$ O/Cu(001) system has been
calculated using locally self-consistent, real space multiple
scattering technique based on first principles. Oxygen atoms are 
found to perturb differentially the long-range Madelung
potentials, and hence the local electronic subbands at neighboring
Cu sites. As a result the hybridization of the oxygen electronic 
states with those of its neighbors leads to bonding of varying 
ionic and covalent mix. Comparison of results with those for the 
$c(2\times2)$ overlayer shows that the perturbation is much stronger 
and the Coulomb lattice energy much higher for it than for the 
$(2\sqrt{2}\times\sqrt{2})R45^{\circ}$ phase. The main driving force
for the 0.5ML oxygen surface structure formation on Cu(001) is thus
the long-range Coulomb interaction which also controls the charge 
transfer and chemical binding in the system.
\end{abstract}


\maketitle

\section{Introduction}

Experimental structural studies of Cu(001) with sub-monolayer
coverage of oxygen shows clear signs of phase instability of
the system (see for example Ref. \onlinecite{kittel}). The recent
studies are a continuation of the debate that has circled around
two main phases: a $c(2\times2)$ structure on an unreconstructed
Cu(001) \cite{kono,holland,dobler,sotto} and a
$(2\sqrt{2}\times\sqrt{2})R45^{\circ}$ structure which involves
a reconstructed surface \cite{wuttig}. Some experiments have
indicated the presence of both phases \cite{tobin,lederer}. Still
other have affirmed the presence of the $c(2\times2)$ overlayer
at low coverages, but only in nanometer size domains and not in
large well-ordered areas \cite{fo96}. Based on these experimental
results, it is now  widely accepted that for coverage lower than
0.34ML, oxygen forms $c(2\times2)$ islands. If the coverage exceeds
this value, the $(2\sqrt{2}\times\sqrt{2})R45^{\circ}$ structure
is formed \cite{kittel}. An array of experimental measurements 
based on surface X-ray diffraction \cite{rob}, low energy electron 
diffraction \cite{atr90,zeng}, surface extended X-ray adsorption 
\cite{lederer}, and photoelectron diffraction \cite{kittel} have 
provided useful, and at times conflicting, information on
$(2\sqrt{2}\times\sqrt{2})R45^{\circ}$  superstructure of O on
Cu(001). One of the common features of the superstructure,
schematically illustrated in Fig. 1, from these studies is the
missing of every fourth (010) row of atoms from the top Cu layer.
Secondly, all studies point to the oxygen atoms occupying sites
located on either side of the missing row between Cu atoms (denoted
as Cu2 in Fig. 1). The structure undergoes further reconstruction
(some displacement of Cu2 along the (100) direction) and relaxation
of mostly the top Cu layer. There is, however, some discrepancy
in the exact nature of the relaxation obtained from different
experiments. For example, according to Ref. \onlinecite{zeng} the
heights $z_O$, $z_{Cu1}$ and $z_{Cu2}$ in Fig. 1, appear in the
order $z_O>z_{Cu2}>z_{Cu1}$, whereas Ref. \onlinecite{kittel}
gives $z_{Cu1}>z_O>z_{Cu2}$ (see Table~\ref{tab:str} for details).

The competition between the two observed phases of O on Cu(001)
and the subsequent reconstruction of the surface with the
$(2\sqrt{2}\times\sqrt{2})R45^{\circ}$ overlayer naturally beg
the question about the origin of this complex behavior. The
driving forces for such structural transformation are expected
to be related to the chemical binding and electronic structure
of the system. Earlier theoretical studies, based on simple
models, have correlated the structural properties of O/Cu(001)
system to a strong $p$O -- $d$Cu hybridization \cite{nors90}
and charge transfer \cite{col} in the vicinity of the surface.
The rationale of these assumptions is supported by calculations,
which indicate the presence of the charge transfer from the metal
atoms to oxygen \cite{madh,bagus} and a strong $p$O -- $d$Cu
hybridization \cite{wiell} for O/Cu(001) $c(2\times2)$ system.
Our recent first principle calculations for the $c(2\times2)$ O
overlayer on Cu(001) \cite{stolb01} also confirm the importance
of considerations of hybridization and charge transfer in the
system and show that these features are eventually
traceable to the long-range Coulomb interaction in the system.
Clearly, the electronic structure and chemical binding in this
system are complex and an adequate description of the
relationship between the electronic and atomic structures of
O/Cu(001) is yet to be established for the
$(2\sqrt{2}\times\sqrt{2})R45^{\circ}$ structure.
Our goal here is to carry out such a study to understand the
driving force for the observed surface reconstruction. For
this purpose we take as benchmarks the two proposed surface
configurations from Ref. \onlinecite{kittel} and
\onlinecite{zeng} and compare their implications.
As we shall see, with minor differences in the electronic
structure, both geometries provide the same rationale for the
ensuing missing row reconstruction. Comparison of the results
with those for the $c(2\times2)$ phase provides further
evidence for the relative stability of the
$(2\sqrt{2}\times\sqrt{2})R45^{\circ}$ O overlayer for 0.5 ML
coverage.

The calculations in the present paper have been carried out using
the real space multiple scattering technique. We have chosen this
particular method because apart from being based on first
principles, it allows calculations of the charge transfer in
the system with ease and provides a good estimate of the nature
of the local bonding. This technique also lends itself nicely to
the calculation of local densities of electronic states, their
projections on the cubic harmonics, and the valence electronic
charge densities. We also calculate the lattice Coulomb energies
and the perturbation of the Madelung potential induced by the 
overlayer.

\section{Computational Details}

Our calculations are based on the density functional theory within
the local density approximation (LDA) \cite{KSh} and multiple
scattering theory. We use the local self-consistent multiple
scattering (LSMS) scheme \cite{Stocks95}, which is designed
for extended heterogeneous solids. Within LSMS a compound is
divided into overlapping clusters called local interaction zones
(LIZs) centered around atoms of different local environments.
For each LIZ, we solve a system of equations for the $T-$ scattering
matrix in the lattice site -- angular momentum representation
for the muffin-tin (MT) potential \cite{Stocks80}:
\begin{equation}
\sum^{}_{n_{1}L_{1}}\left\{ \delta ^{n^\prime n_1}_{L^\prime L_1}
t^{-1}_{n^\prime l^\prime} - g^{n^\prime n_1}_{L^\prime L_1} \right\}
T^{n_1 n}_{L_1 L} = \delta_{LL^\prime}^{nn^\prime } ,
\end{equation}
\noindent where $n$ denotes the lattice site number, $L=l,m$ are
angular momentums, $t_{nl}$ are the single-site scattering matrix
elements, $g_{LL_1}^{nn_1}$ denote the free-electron Green's
function matrix elements, and $T_{LL^{\prime }}^{nn^{\prime }}$
are scattering path matrix elements. The solutions are used to
determine the LIZ Green's functions, the local densities
$N_l^n(E)$ of electronic states, their projections
$N_{lm}^n(E)$ on the cubic harmonics, and local valence charge
densities for the atoms located at the center of each LIZ.
These charge densities are used further to solve the Poisson's
equation for the entire extended system and hence to build a new
potential for the next iteration of the self-consistent process.
An application of this method to materials related to the work
here has demonstrated its high efficiency and reliability
\cite{stolb97,stolb99,stolb01}.

To take into account surface effects on the MT-potential, we follow
the approach developed in Ref. \onlinecite{ujf}. In this approach
the space in the vicinity of the surface is divided into layers
belonging to different atomic planes parallel to the surface and
the interstitial charge density $\rho_{i}$ is supposed to be
layer-dependent reflecting the asymmetry of the system. The detailed
description of the technique is given elsewhere \cite{stolb01}. Here
we only remind our readers that in our calculations the Madelung
potential includes monopole $M_{i\alpha,j\beta}^{00}$, dipole
$M_{i\alpha,j\beta}^{10}$ and interstitial $V_{ij}[\rho_j]$
terms\cite{ujf}:

\begin{equation}
V_{i\alpha}^{Mad}=\sum_{j,\beta} \left ( q_{j\beta}M_{i\alpha,
j\beta}^{00}+d_{j\beta}M_{i\alpha,j\beta}^{10}+V_{ij}[\rho_j]
\right ),
\end{equation}
where $i,j$ denote the layer number, while $\alpha, \beta$ number the
atoms in the 2D unit cell, $d_{j\beta}$ is a dipole moment of the
MT charge density, and the effective charge

\begin{equation}
q_{j\beta}=Z_{j\beta}-q_{j\beta}^{MT}+\frac{4\pi}{3}\rho_jR_{j\beta}^3.
\end{equation}

The exchange and correlation parts of the potential are determined
within LDA using the technique described in Ref. \onlinecite{GunLun}.

In this work the above formalism is applied to the case of
$(2\sqrt{2}\times\sqrt{2})R45^{\circ}$ O overlayer on Cu(001).
Fig. 1 is a schematic representation of the system. The parameters
for the two proposed configurations from Ref. \onlinecite{kittel}
and \onlinecite{zeng}, hereafter referred as "structure I" and
"structure II", are summarized in Table~\ref{tab:str}. The second
and lower Cu layers are assumed to occupy bulk-like positions.
The presence of oxygen atoms and the missing rows reduce the symmetry
of the system and create nonequivalent environments for Cu atoms
belonging to the same layer. We take this into account by allowing
atoms of the first three layers to be nonequivalent as needed.
They are marked by different numbers in Fig. 1. Thus we build LIZ's
around oxygen and nine nonequivalent Cu atoms within the top five
layers. We take the sites of missing Cu to be occupied by vacancies
represented by MT-spheres with zero core electron density and
requiring an additional LIZ to be built around the vacancy. These
LIZ's contained 62 to 91 atoms depending on the local configuration,
and the requirement that the calculated characteristics of the bulk
system closely match those obtained from other reliable methods in
the literature.

\section{Results and Discussion}

Adsorption of gases on surfaces causes both short- and long-range
effects on the electronic structure of the system. The short-range
effect originates from changes in the local environment of some
surface atoms which lead to modifications of their chemical bonds.
The long-range effect induced by the Coulomb interaction between
the chemisorbed overlayer an the substrate, on the other hand, may
cause differing shifts in the local potentials at atomic sites with
differing environments. As a result, charge transfer may be induced
and the electronic state hybridizations at these sites may be
modified.

Here we examine in details both the short- and long-range effects of
the chemisorption of an oxygen overlayer on Cu(001) with the view of
understanding the driving forces responsible for the observed missing
row reconstructed $(2\sqrt{2}\times\sqrt{2})R45^{\circ}$
superstructure. We present first a comparative study of the
short-range effect in structures I and II through examination of the
local densities of electronic states at the oxygen and top Cu sites
and their hybridizations. To gain further insight into the nature
of the O-Cu bonding and subtle differences emerging from
structures I and II, we resort to an analysis of the projections of
these densities of states on cubic harmonics. These
short-range effects, however, are related to the long-range ones
responsible for shifts in the local potentials at each site. We do
this next through a systematic evaluation of the effects of the
Coulomb interaction on local potentials. This is followed by an
analysis of the charge transfer in the system. Finally, we comment
on the relative stability of the $c(2\times2)$ phase, in view of the 
findings here for the $(2\sqrt{2}\times\sqrt{2})R45^{\circ}$ phase.

\subsection{Local densities of the pO- and dCu-electronic states}

The calculated densities of the $p$-states of the oxygen and
$d$-states of its nearest neighbors Cu1, Cu2 and Cu3 atoms for 
structures I and II are plotted in Fig. 2. The densities of the
$d$Cu4 through $d$Cu9 states ($N_d^{Cu}(E)$) are shown in Fig. 3.
Since the difference between $N_d^{Cu}(E)$ of Cu6 and Cu7 is 
found to be negligible, only one is displayed. Figs. 2 and 3
indicate a dramatic difference in $N_l^n(E)$ for atomic sites
with different environments. The $N_d^{Cu}(E)$ of the Cu6 and Cu8
atoms belonging to the third and fourth layers, respectively,
differ slightly from bulk values, while that of Cu9 coincides
with that of the bulk. The $d$Cu4- and $d$Cu5-states, on the other
hand, consist of a pronounced narrow peak together with some low
energy structure. Finally, the densities of the $p$O-states
and $d$-states of oxygen's nearest neighbors are found to
be significantly split and structured. This splitting results in
two substructures ("a" and "b" in Fig. 2), which have the same
energetic position and similar shape for all spectra. Such a
behavior clearly indicates a strong hybridization and covalent
binding between the $p$O-states and the $d$-states of the
neighboring Cu atoms. A relatively high intensity of the low
energy peaks "a" in $N_d^{Cu3}(E)$ suggests that the
$d$Cu3-states are more involved in the $p-d$ hybridization than
the $d$Cu1- and $d$Cu2-states. To understand the origin of this
difference we should remember that the degree of
hybridization of electronic states  depends on their spatial
and energetic overlap. Since, in the present structures, the
O-Cu3 interatomic distance is longer than O-Cu1 and O-Cu2 ones
\cite{zeng,kittel}, the strong $p$O-$d$Cu3 hybridization does
not result from the larger spatial overlap. The energetic overlap,
on the other hand, can be quantitatively characterized within the
multiple scattering theory by comparison of the single-site
resonance energies ($E_{res}^n$) for neighboring sites, which
reflect energetics of the local potentials and primarily govern
the energy of local subbands \cite{wein}. The calculated
$E_{res}^n$ for structures I and II, in Table~\ref{tab:res}, show
substantial variation of $E_{res}^n$ from site to site. As we shall
see in section C, these variations are a result of the long-range
interaction and charge transfer in the system. Since from
Table~\ref{tab:res}, the energetic separation between the $p$O and
$d$Cu3 resonances is smallest, we can conclude that the strong
covalent $p$O -- $d$Cu3 binding is caused by a large energetic
overlap of these states. Also, as $E_{res}^{Cu4}$ and $E_{res}^{Cu5}$
are much higher than the others in Table~\ref{tab:res}, there is only
a small coupling of the $d$Cu4- and $d$Cu5-states with those of
their neighbors as indicated by the low intensity of the structures
seen in the low energy region in Fig. 3.

Having established the strong coupling between the $p$O and $d$Cu3 
electronic states, we proceed next with an analysis
of the projections of the local densities of states on cubic 
harmonics ($N_{lm}^n(E)$) which provide more detailed description of 
covalent binding in the system and help isolate the reasons for the
a noticeable difference between the shape and splitting of for 
structures I and II. 

\subsection{Projections of the densities of states on the cubic
harmonics}

Since in the geometric structures under consideration, the local
environment of the oxygen atoms have a reduced symmetry (see Fig. 4)
because of the missing rows, the local electronic states are expected
to be anisotropic. Therefore, to build a consistent picture of
chemical binding in the system, it is important to analyze the
projections of electronic state $N_{lm}^n(E)$ on the cubic harmonics.
Plots in Fig. 5 of such projections of $p$O- and the five 
$d$Cu3-states, calculated for structure I, indicate that they differ 
dramatically from each other. The O-$p_z$ and Cu3-$d_{3z^2-1}$ states 
are found to be strongly split. They form a common subband 
represented by peaks "a" and "b", which are separated by about 
0.5Ry. This strong splitting together with the involvement of major 
part of these states in hybridization (redistributed into "a"
and "b" regions) are a clear signature of strong covalent binding. 
The O-$p_x$ and Cu3-$d_{xz}$ as well as O-$p_y$ and Cu3-$d_{yz}$ 
states also form common split subbands, but not as strongly as the 
O-$p_z$ -- Cu3-$d_{3z^2-1}$ states. The cause of this difference in 
the hybridization becomes clear through a comparison of the angular 
distribution of corresponding cubic harmonics, which shows that the
O-$p_x$ -- Cu3-$d_{zx}$ and O-$p_y$ -- Cu3-$d_{zy}$ spatial overlap
is lower than that of O-$p_z$ -- Cu3-$d_{3z^2-1}$ (Fig. 5). The 
Cu3-$d_{xy}$ and Cu3-$d_{x^2-y^2}$ states cannot be coupled with 
the oxygen electronic states by symmetry. Instead they hybridize 
weakly with the electronic states of Cu4 and Cu5, as reflected by 
low intensity structure at high energies in Fig. 5. The weakness
of this particular hybridization can be traced to the high energetic 
separation of the single-site resonances (see Table~\ref{tab:res}).

In Fig. 6, the corresponding splitting of the
O-$p_z$ -- Cu3-$d_{3z^2-1}$ subband for structure II is found to be
larger (5.7 Ry versus 5.0 Ry) than that for structure I, implying
stronger covalent binding. Although for the O-$p_x$ -- Cu3-$d_{zx}$ 
and O-$p_y$ -- Cu3-$d_{zy}$ subbands it is harder to quantify the 
splitting, Figs. 5 and 6 indicate that these subbands are 
also split stronger for structure II than for structure I. Again, 
these differences in splitting can be traced to the energetic separation 
between the $p$O and $d$Cu3 single-site resonances found for the two
structures (see Table~\ref{tab:res}).

The projections $N_{dm}^{Cu}(E)$ calculated for the other nearest
neighbors of oxygen, namely, Cu1 and Cu2 are shown in Fig. 7 and
8, respectively. Only a minor part of the $d$Cu1- and
$d$Cu2-electronic states are found to contribute to the peaks
responsible for the $p$O -- $d$Cu hybridization. As mentioned above,
the single-site $d$Cu1- and $d$Cu2-resonances are energetically
separated from the $p$O-resonance that reduces the hybridization. The
low energetic structures of $N_{dm}^{Cu1}(E)$ and $N_{dm}^{Cu2}(E)$
align with "a" peaks of $N_{pm}^O(E)$, which are formed with the
$p$O-$d$Cu3 hybridization. This suggests that the $d$Cu1- and
$d$Cu2-states just admix with the $p$O-$d$Cu3-subband and the O-Cu1
and O-Cu2 covalent bonds are weaker than the O-Cu3 bond.

\subsection{Long-range Coulomb interaction and charge transfer}

For a deeper understanding of the factors controlling the
electronic and atomic structures of the O/Cu(001) system, we turn
now to considerations of long-range effect of the Coulomb
interaction as expressed by the Madelung potential and the subsequent
charge transfer in the surface region. To get a simple and intuitive
picture we first calculate the Madelung potential for several surface
structures of Cu(001) with 0.5 ML of O, within a simple "frozen" 
charge density approximation using Eq. 2. Namely, we calculate
the effective charges $q_{i \alpha}$ introduced in Eq. 3  using
bulk-like electronic density for Cu and atom-like density for O.
Such an approach gives us a pure geometric effect of the long-range
interaction. As expected, the presence of the surface itself causes
only  small perturbation in the Madelung potential. This is
represented by the low values in the column under "clean" (Cu(001) in
Table~\ref{tab:Mad1}. The admission of 0.5 ML oxygen on bulk
terminated Cu(001) causes large perturbation in the Madelung potential
at atomic sites in the top few layers, as seen in the column denoted
by BT in Table~\ref{tab:Mad1}. By removing every fourth row of Cu atom
from this bulk terminated O/Cu(001) system, we find the perturbation
of the Madelung potential to be significantly reduced in the top Cu
layer and slightly increased in some of the Cu sites below (see column
labeled MR in Table~\ref{tab:Mad1}). The structures BT and MR are
naturally artificial not only in terms of the positions of the top
layer Cu atoms, but also that of the O which is taken to occupy the
hollow sites in the surface plane. They have been chosen to provide
benchmarks for the relative stability of the two observed O
superstructures. The calculated perturbation of the Madelung 
potentials for structure II, i.e. 
$(2\sqrt{2}\times\sqrt{2})R45^{\circ}$ O phase with Cu atoms in 
relaxed and reconstructed positions, in Table~\ref{tab:Mad1}, 
indeed display significant reductions from the values of BT and MR 
configurations, for all atomic sites. Finally, in Table~\ref{tab:Mad1} 
we have also included the calculated quantities for the $c(2\times2)$ 
structure (parameters taken from Ref. \onlinecite{kittel} for a coverage
less than 0.34 ML) for comparison. Clearly, the $c(2\times2)$ phase,
albeit for 0.5 ML coverage, invokes a much stronger perturbation
of the Madelung potential than the one in structure II.

An electronic response to the strong perturbation of the Madelung
potential is expected to be charge transfer in the vicinity of the
surface. The resulting charge distribution would then determine the
degree of ionicity of the system, and would influence single-site
resonance energies, and consequently, the formation of covalent bonds.
Charge transfer is thus a very important characteristic of systems.
To characterize it quantitatively, the system has to be divided into
spaces belonging to each atom and the charge within this space
calculated. In the framework of the MT-approximation, non-overlapping
spheres centered on the atomic sites can naturally represent these
spaces. A complexity of the present system is that interatomic bond
lengths for a number of atoms are different because of the differences
in their local environment. Thus different volumes belong to different
atomic sites. To make the site charges comparable, we have built
the spheres with the fixed radius equal to one-half of the shortest
interatomic distance (1.807 \AA) and calculated the local valence
charges by integrating the self-consistent valence electronic
density over these spheres. The results in Table~\ref{tab:char}
show a significant electronic charge transfer from Cu to O in both
structures I and II. The charge transfer pattern is, however,
interesting as Cu2, Cu4 and Cu5 donate part of their charge but
Cu1 and Cu3 do not. These latter two nearest neighbors of O, in fact, 
gain a small amount. Such a complex character of the charge transfer 
reflects the local potential perturbations caused by the missing Cu 
rows induced by the oxygen overlayer. Indeed, from Table~\ref{tab:Mad1} 
and Table~\ref{tab:char} it follows that sites loose
electronic charge if the initial Madelung potential is higher and
gain it if the potential is lower. Table~\ref{tab:char} also reflects 
subtle differences in the charge transfer to the different sites in the 
two structures (I and II) considered here. Recall that covalent bonds 
have also been found stronger in structure II as compared to structure I.

Naturally the Coulomb interaction of the different sublattices plays
a very important role in the above considerations. To get further insights
into the difference arising from the two structural models chosen in
this paper, we compare the Coulomb lattice energies $E_{lat}$ for
these O/Cu(001) phases. Within the MT approximation \cite{schmidt}
$E_{lat}$ can be expressed as follows:
\begin{equation}
 E_{lat}= \frac{1}{2}\sum_{i,\alpha }\sum_{j,\beta }
M_{i\alpha,j\beta }q_{i\alpha }q_{j\beta } +
\sum_{i,\alpha} \frac{1}{r_{MT}^{i\alpha }}
(1.2(\omega_{i\alpha }\rho_i)^2 +
3q_{i\alpha }\rho_i\omega_{i\alpha}),
\end{equation}
where $r_{MT}^{i\alpha}$ and $\omega_{i\alpha}$ are the MT-radii and
MT-sphere volumes, respectively. The first sum of Eq. 4 represents the
Madelung energy and the second comes from the charge neutrality
condition. The $E_{lat}$ thus includes the Coulomb interaction
between the MT charges and that between them and the layer-dependent
uniform interstitial charge density in the system. For comparative
purposes it is sufficient to include the contribution of the oxygen
overlayer and only the top five Cu layers to $E_{lat}$. Eq. 4 then
yields values of -13.864Ry and -15.420Ry for $E_{lat}$, per 2D unit
cell, for structure I and structure II, respectively. The lower value 
of $E_{lat}$ along with the finding of stronger O-Cu covalent 
binding in structure II than in structure I, indicate preference
for the former structure.

We have also calculated the Coulomb lattice energy for the
$c(2\times2)$ O overlayer on Cu(001) using our self-consistent
calculation results from Ref. \onlinecite{stolb01} and obtained the
value of $E_{lat}=$-8.582Ry. This is much higher than that in the
case of both structures I and II. This result is not surprising as
we have already shown in Table~\ref{tab:Mad1} that the perturbation
to the local Madelung potential at the atomic sites is also larger
for the $c(2\times2)$ superstructure. Based on these findings we
conclude that the long-range Coulomb interaction is the main factor
which controls the phase formation in the O/Cu(001) system and makes
the $(2\sqrt{2}\times\sqrt{2})R45^{\circ}$ phase preferable at 
oxygen coverage of 0.5ML in agreement with experimental findings.

\section{Conclusions}

The results presented above, provide the following microscopic picture
for the formation of the $(2\sqrt{2}\times\sqrt{2})R45^{\circ}$ O
overlayer on Cu(001). Oxygen adsorption in the four-fold hollow sites
of Cu(001) induces a strong Coulomb perturbation on the top layer Cu
sites. Missing row reconstruction of the Cu surface, lattice relaxation,
and charge transfer help significantly reduce this perturbation. In
response to the Coulomb perturbation, the electronic charge transfer
from the host surface to O is found to have a complex character,
involving two of the four nearest neighbors and two next nearest
neighbors of O. The perturbation is screened partially and the
self-consistent local potentials are such that the energetic separation
between the single site $p$O- and $d$Cu-resonances is different for
different Cu sites. This energetic separation essentially determines
the degree of the $p$O-$d$Cu hybridization and the strengths of the
covalent bonds formed between oxygen and its nearest neighbors. Thus,
the Cu3 atom located right under O is involved in strong covalent
binding with O, whereas weak covalent bonds are formed between the
O and its other nearest neighbors. The O-Cu3 binding is found to
be stronger in structure II than in structure I. Furthermore, as
the Coulomb lattice energy is lower for the structure II than for
structure I. In extending these consideration to the observed
$c(2\times2)$ O overlayer on Cu(001) (for coverage less than 0.34ML),
we conclude that the $c(2\times2)$ structure is unstable, for half
monolayer coverage, because it induces a strong perturbation at the
surface that increases the Coulomb lattice energy of the system. Thus,
according to our results, the long-range Coulomb interaction is the
main driving force for the O/Cu(001) surface atomic structure
formation.

Finally, we note that adsorption processes on metal surfaces are
extensively studied by means of such methods as Monte Carlo or
molecular dynamics simulations. Many of them are based on the model
potentials, which depend only on the local atomic environment, namely
on the interatomic distance and the number of the nearest neighbors.
Our findings indicate that an application of such approaches to
O/Cu(001) can hardly be successful. Indeed, according to our results,
the long-range interaction in the system governs not only charge
transfer (as usual), but also such "short-range" effects as the
electronic state hybridization and covalent bond formation. Such an
implicit impact cannot be taken into account even if a model potential
somehow includes the long-range interaction along with a traditional
short-range part. This suggests that a proper model potential for
O/Cu(001) can be built either by parameterization of the energetic
surface obtained from first principle calculations or by development
a sophisticated form, which takes into account the interplay between
the long- and short-range parts.

\begin{acknowledgements}

This work was supported by the National Science Foundation
(Grant No CHE9812397). The calculations were performed on the
Origin 2000 supercomputer at the National Center for
Supercomputing Applications, University of Illinois at
Urbana-Campaign under Grant No DMR010001N. The work of TSR
was also facilitated by the award of an Alexander von
Humboldt Forschungspreis.
\end{acknowledgements}

\newpage

{\bf Figure Captions}

\noindent Fig. 1. Schematic illustration of the O/Cu(001)
$(2\sqrt{2}\times\sqrt{2})R45^{\circ}$ structure.

\noindent Fig. 2. Densities of the $p$-electronic states of oxygen and $d$-
electronic states of its nearest neighbors calculated for
structure I (solid line) and structure II (dashed line).

\noindent Fig. 3. Densities of the $d$-electronic states of the Cu4 through Cu9
atoms calculated for structure I (solid line) and structure II
(dashed line).

\noindent Fig. 4. Local environment of oxygen atom in the 
$(2\sqrt{2}\times\sqrt{2})R45^{\circ}$ O/Cu(001) structure.

\noindent Fig. 5. Projections of the $p$O- and $d$Cu3-electronic states
calculated for structure I.

\noindent Fig. 6. Projections of the $p$O- and $d$Cu3-electronic states
calculated for structure II.

\noindent Fig. 7. Projections of the $d$Cu1-electronic states calculated
for structure I (solid line) and the structure II (dashed line).

\noindent Fig. 8. Projections of the $d$Cu2-electronic states calculated
for structure I (solid line) and structure II (dashed line).
\newpage
\begin{table}
\caption{\label{tab:str} Structural parameters proposed from Ref.
\cite{kittel} (structure I) and Ref. \cite{zeng} (structure II)
which are used in our calculations}
\begin{ruledtabular}
\begin{tabular}{ccc}
Parameter & Structure I & Structure II \\
\hline
$z_O$ (\AA) & 2.05 & 2.14 \\
$z_{Cu1}$(\AA) & 2.14 & 1.94 \\
$z_{Cu2}$(\AA) & 1.88 & 2.04 \\
$\delta x_{Cu2}$(\AA)\footnotemark[1] & 0.29 & 0.30 \\ %
\end{tabular}
\end{ruledtabular}
\footnotetext[1]{the Cu2 atom displacement along the x direction
toward the missing row}
\end{table}
\begin{table}
\caption{\label{tab:res}Energetic positions (Ry) of the single site
$p$O- and $d$Cu-resonances $E_{res}^n$ counted from the MT-zero level}
\begin{ruledtabular}
\begin{tabular}{ccc}
Site & Structure I & Structure II \\
\hline
 O & 0.232 & 0.222 \\
Cu1 & 0.354 & 0.377 \\
Cu2 & 0.416 & 0.447 \\
Cu3 & 0.316 & 0.283 \\
Cu4 & 0.492 & 0.493 \\
Cu5 & 0.533 & 0.516 \\
Cu6 & 0.403 & 0.408 \\
Cu7 & 0.408 & 0.417 \\
Cu8 & 0.382 & 0.388 \\
Cu9 & 0.382 & 0.382 \\
\end{tabular}
\end{ruledtabular}
\end{table}
\begin{table}
\caption{\label{tab:Mad1} The variation of the Madelung potential
(in Ry) at different atomic sites with respect to the bulk value
for the following configurations: clean Cu(001) surface (Clean);
O on bulk terminated Cu(001) (BT)); $(2\sqrt{2}\times\sqrt{2})R45^{\circ}$
 O on bulk terminated O/Cu(001) with missing rows (MR); structure II;
$c(2\times2)$ O on Cu(001) ($c(2\times2)$)}
\begin{ruledtabular}
\begin{tabular}{ccccccc}
Layer & Site & Clean & BT & MR & Structure II & $c(2\times2)$ \\
\hline
 & O & - & -0.329 & -0.511 & -0.880 &  -0.379 \\
 I & Cu1 & -0.007 & 1.438 & 0.495 & 0.219 & 1.810 \\
 & Cu2 & -0.007 & 1.432 & 1.022 & 0.660  & 1.810 \\
\hline
 & Cu3 & 0.002 & 0.149 & -0.207 & 0.008  & -0.113 \\
 II & Cu4 & 0.002 & 0.285 & 0.271 & 0.299 & -0.113  \\
 & Cu5 & 0.002 & 0.149 & 0.478 & 0.162  &  -0.113  \\
\hline
 III & Cu6 & 0.001 & 0.020 & 0.016 & 0.011 &  0.01 \\
 & Cu7 & 0.001 & 0.014 & 0.100 & 0.037 & 0.01  \\
\hline
 IV & Cu8 & 0.0 & 0.003 & 0.003 & 0.0  & 0.0 \\
\hline
 V & Cu9 & 0.0 & 0.004 & 0.0 & 0.0 & 0.0
\end{tabular}
\end{ruledtabular}
\end{table}
\begin{table}
\caption{\label{tab:char} Variations of the local valence electronic
charge from the bulk-like charge for Cu sites and from the
atomic-like charge for the O site.}
\begin{ruledtabular}
\begin{tabular}{cccc}
Layer & Site & Structure I & Structure II \\
\hline
 & O & 0.518 & 0.508 \\
 I & Cu1 & 0.102 & 0.015 \\
 & Cu2 & -0.131 & -0.287 \\
\hline
 & Cu3 & 0.069 & 0.050 \\
 II & Cu4 & -0.164 & -0.175 \\
 & Cu5 & -0.263 & -0.361 \\
\hline
III & Cu6 & 0.006 & -0.017 \\
 & Cu7 & -0.005 & -0.016 \\
\hline
IV & Cu8 & 0.013 & 0.014 \\
\hline
V & Cu9 & 0.0 & 0.0 \\
\end{tabular}
\end{ruledtabular}
\end{table}

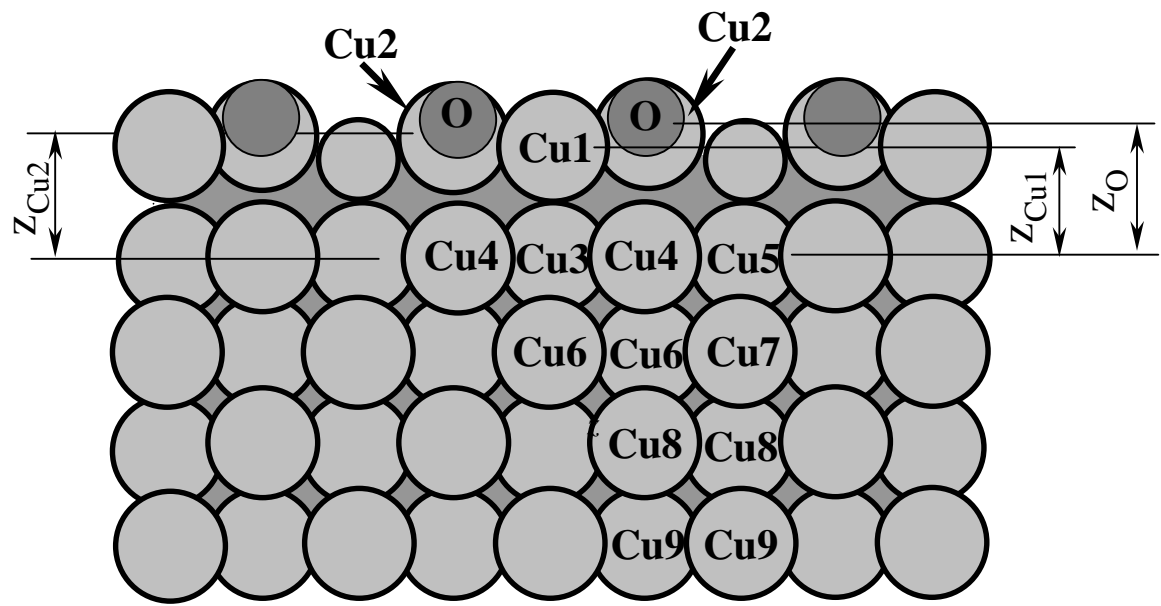
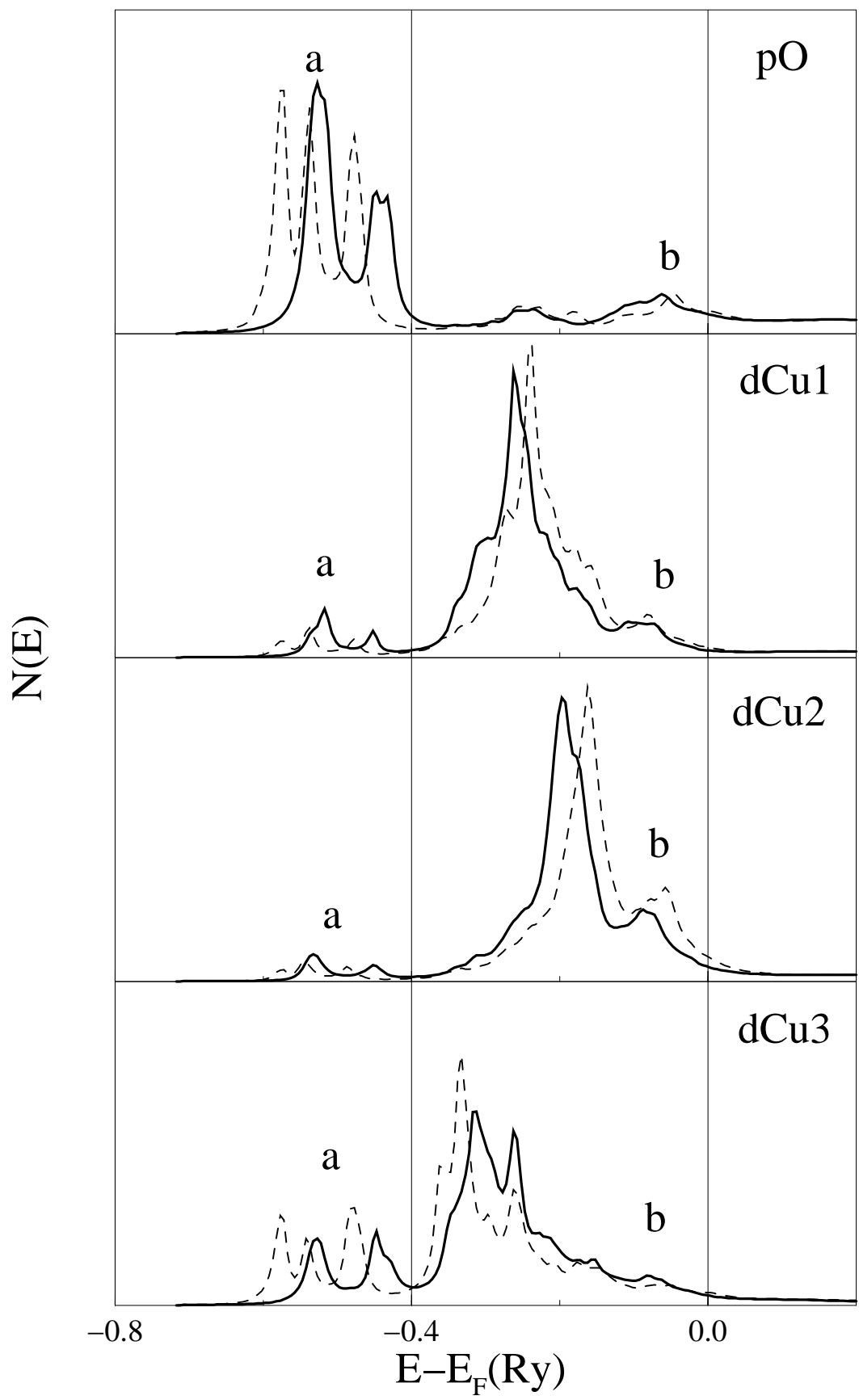
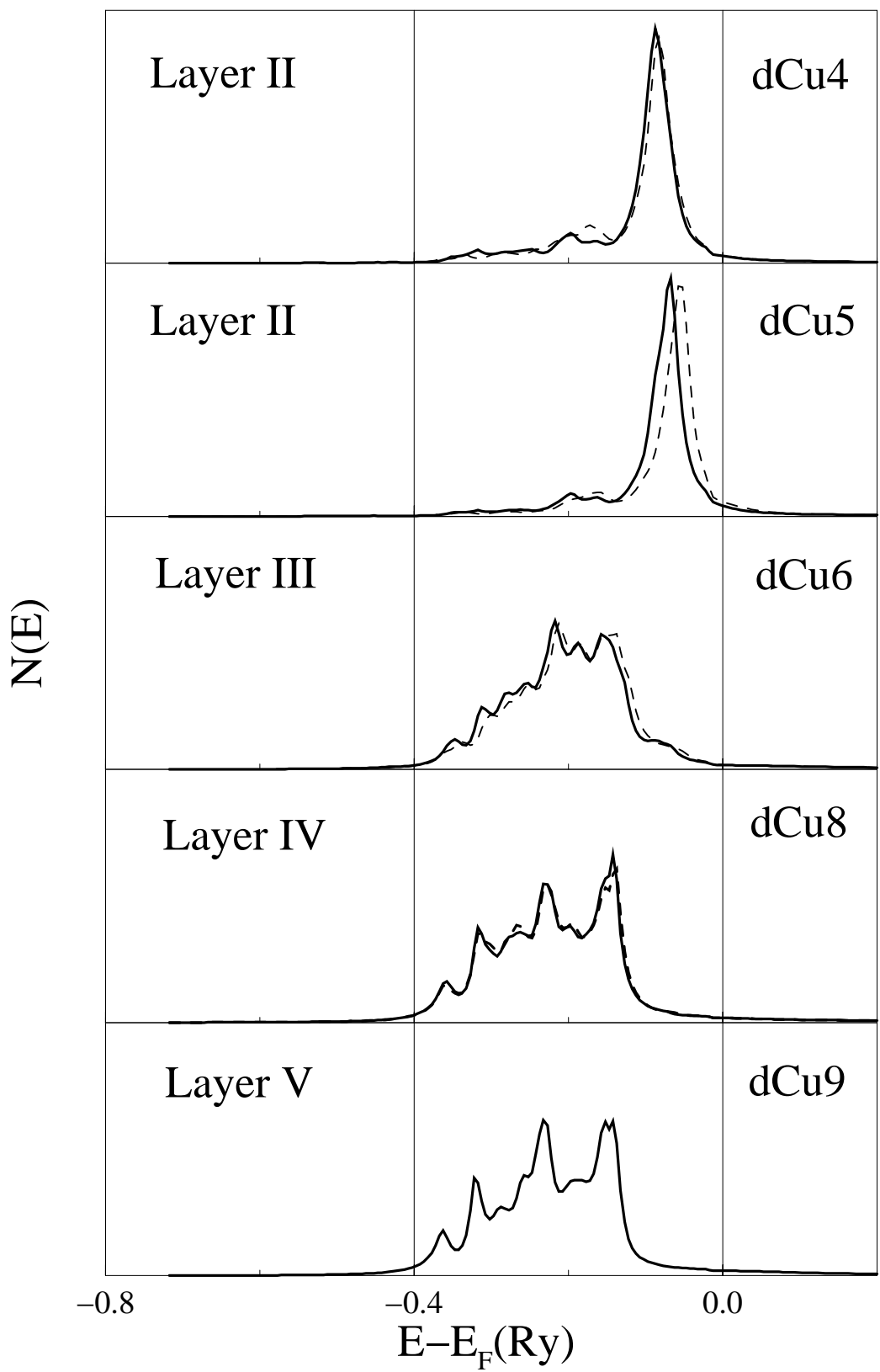
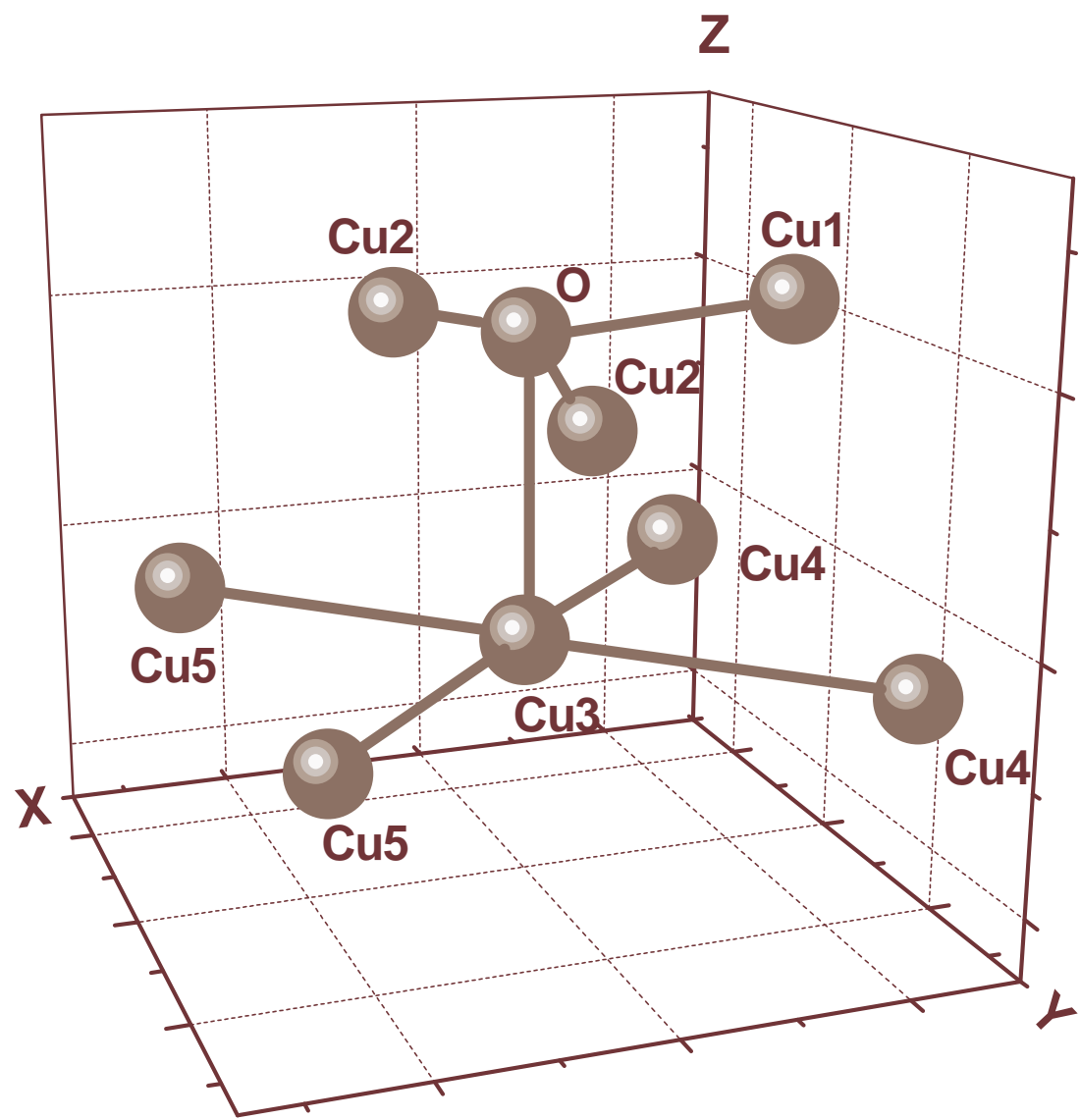
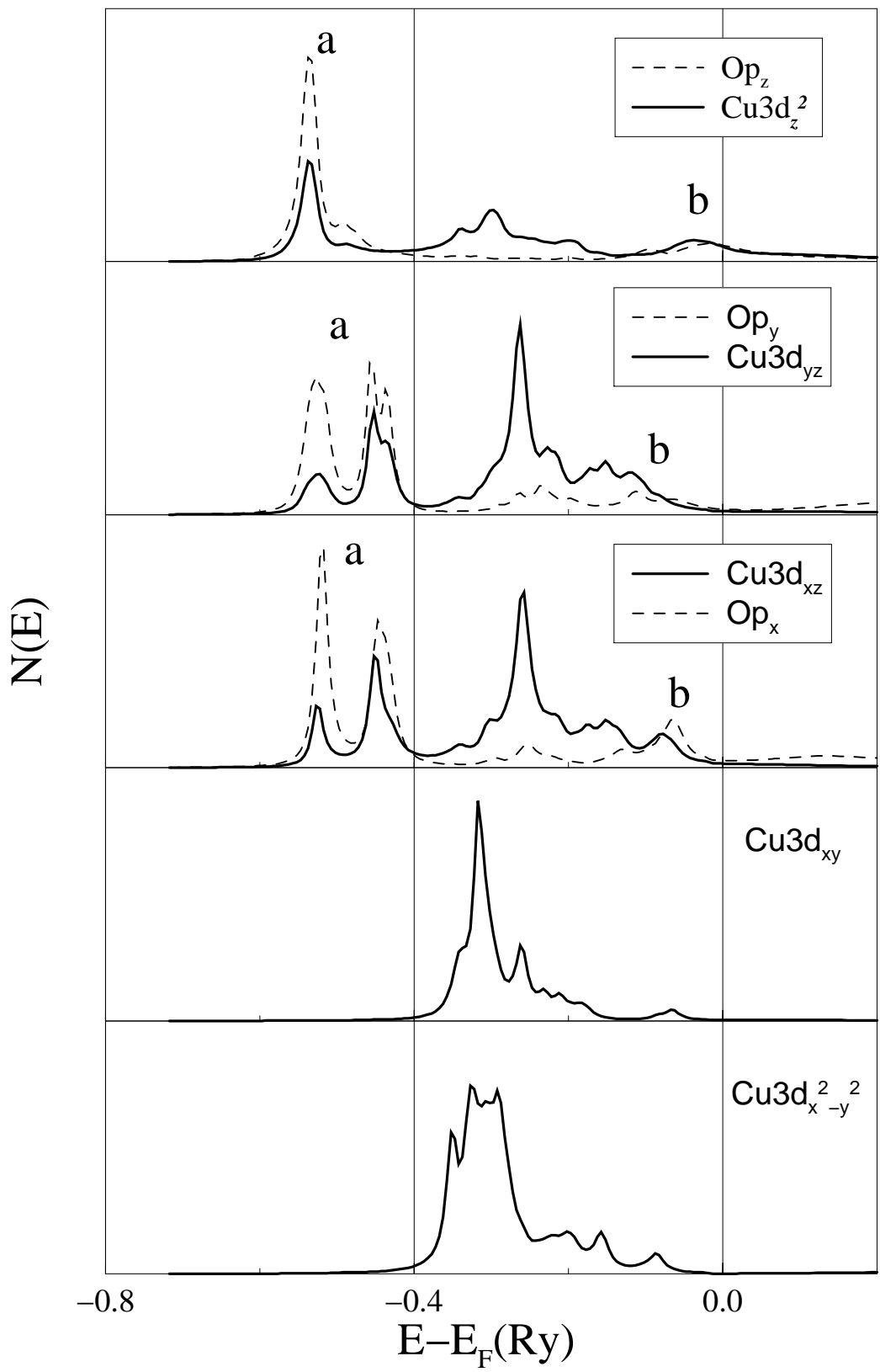
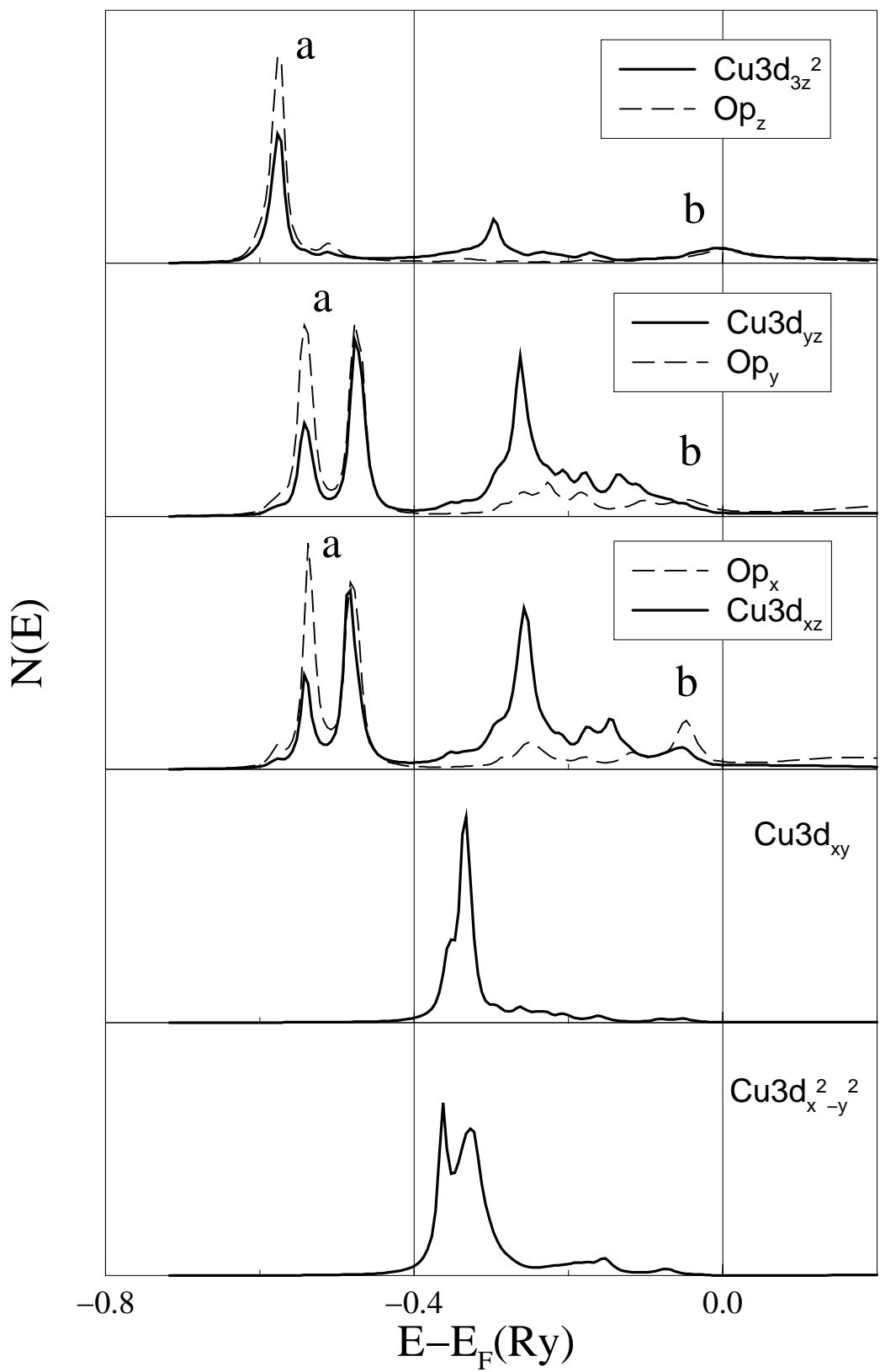
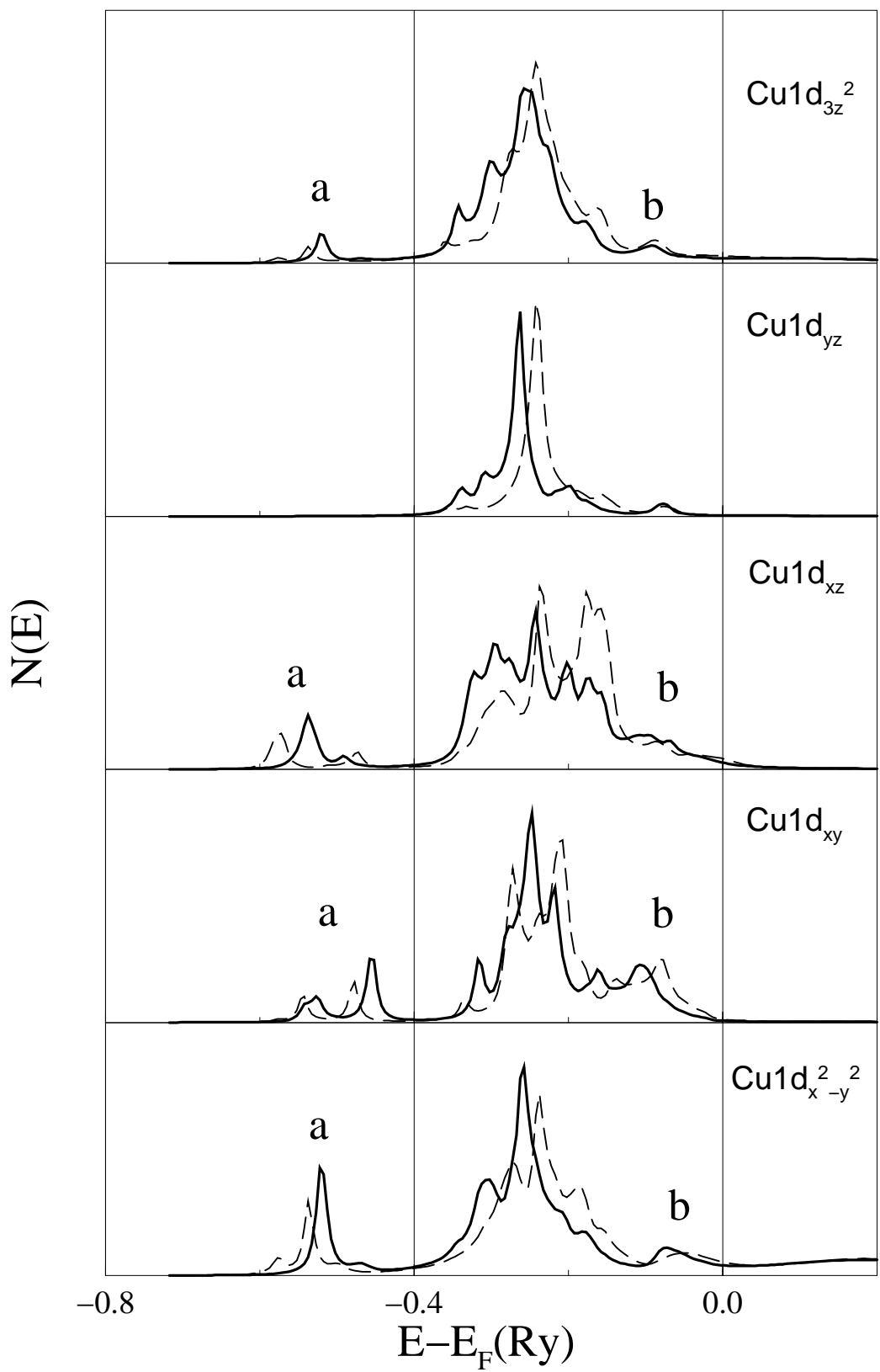
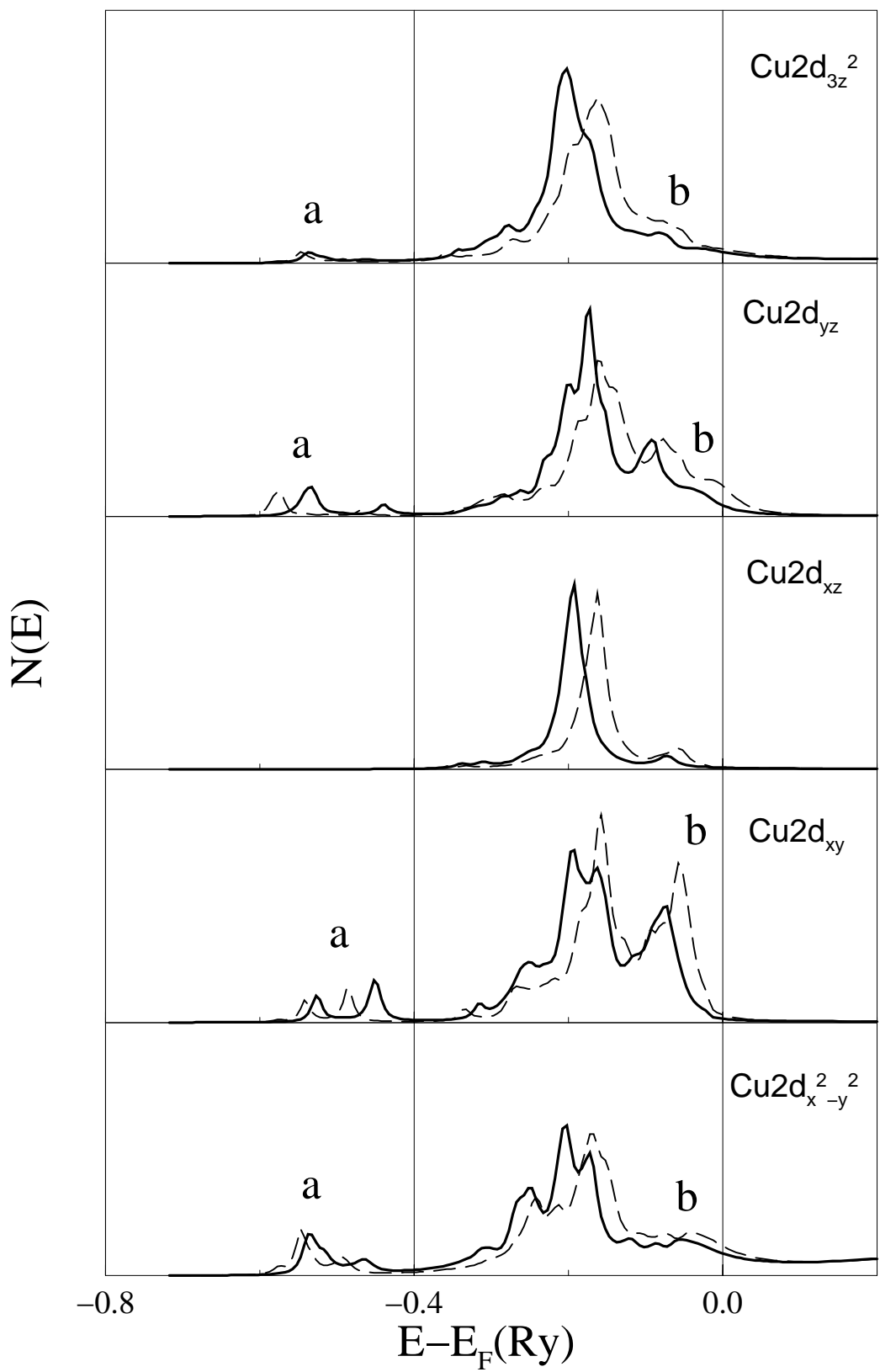
\end{document}